\documentclass[twocolumn,epjc3]{svjour3}
\smartqed  
\RequirePackage{graphicx}
\begin{document}

\title{Constraining axion-nucleon coupling constants
 from measurements of effective Casimir pressure by means
 of micromachined oscillator}

\titlerunning{Constraining axion-nucleon coupling constants}

\author{V.~B.~Bezerra\thanksref{addr1} \and
G.~L.~Klimchitskaya\thanksref{addr2,addr3,addr1}\and
V.~M.~Mostepanenko\thanksref{e1,addr2,addr3,addr1}\and
C.~Romero\thanksref{addr1}
}                     
\authorrunning{V.~B.~Bezerra et al.}
\thankstext{e1}{e-mail: vmostepa@gmail.com}
\institute{Department of Physics, Federal University of Para\'{\i}ba,
C.P.5008, CEP 58059--970, Jo\~{a}o Pessoa, Pb-Brazil \label{addr1}
\and
Central Astronomical Observatory
at Pulkovo of the Russian Academy of Sciences,
St.Petersburg, 196140, Russia \label{addr2}
\and
Institute of Physics, Nanotechnology and
Telecommunications, St.Petersburg State
Polytechnical University, St.Petersburg, 195251, Russia
\label{addr3}
}
\date{Received: date / Revised version: date}
%
\maketitle

\abstract{
Stronger constraints on the pseudoscalar coupling constants
of an axion to a proton and a neutron are obtained from an
indirect measurement of the effective Casimir pressure
between two Au-coated plates by means of micromechanical
torsional oscillator. For this purpose, the additional
effective pressure due to two-axion exchange is calculated.
The role of boundary effects and the validity region of the
proximity force approximation in application to forces of
axion origin are determined. The obtained constraints are
up to factors of 380 and 3.2 stronger than those found
recently from other laboratory experiments and are
relevant to axion masses from $10^{-3}\,$eV to 15\,eV.
%
} 
\section{Introduction}

Starting from the prediction of axions in 1978 \cite{1,2},
axion physics has become a wide subject stimulating
development of elementary particle theory, gravitation and
cosmology (see \cite{3,4} for a review).
Axions are pseudoscalar particles which appear as a
consequence of breaking the Peccei and Quinn symmetry \cite{5}.
They provide an elegant solution for the problem of strong
{\it CP} violation and large electric dipole moment for the neutron
in QCD. Since the proper QCD axions were
constrained to a narrow band in parameter space \cite{5a},
a lot of invisible axion-like particles
have been proposed in different unification schemes.
Among others, the models of the {\it hadronic} (KSVZ) \cite{6,7}
and the  GUT (DFSZ) \cite{8,9} axions, which can be used to
solve the problem of strong
{\it CP} violation in QCD, have attracted particular attention
(see, for instanse, a number of variants of the model of hadronic
axion containing the relationship between the axion-nucleon
coupling constant and the Peccei-Quinn symmetry breaking
scale \cite{9a,9b}).
At the moment axion-like particles with masses $m_a$ from
approximately $10^{-5}\,$eV to $10^{-2}\,$eV and of about
1\,MeV are not excluded by astrophysical constraints \cite{4,10}.
Keeping in mind that the latter may be more model-dependent
than the laboratory constraints \cite{11,12}, it seems
warranted to look for some alternative phenomena which could be
used for constraining axion-like particles of any mass.
Additional interest in this subject is due to the role
of axions as possible constituents of dark matter \cite{12a,12b}.

Previously constraints on axion-nucleon coupling constants have
been obtained \cite{13,14,15} from the laboratory experiments
of E\"{o}tvos \cite{15,16} and Cavendish \cite{17,21a} type.
At first, this analysis was performed for massless axions but
later it was generalized \cite{18} for the case of massive ones.
The resulting constraints were found in the range of axion masses
from approximately $10^{-9}\,$eV to $10^{-5}\,$eV.
In \cite{19} constraints on the axion-nucleon coupling constants
were obtained from measurements of the thermal Casimir-Polder
force between a Bose-Einstein condensate of ${}^{87}$Rb atoms
and a SiO${}_2$ plate \cite{48}. These constraints refer to
larger axion masses from $10^{-4}\,$eV to 0.3\,eV. In fact,
the effective potential arising between two fermions from the
exchange of a pseudoscalar axion-like particle is spin-dependent
\cite{18}. Taking into account that the test bodies in the
experiments \cite{15,16,17,21a,48} are unpolarized, the aditional
force constrained in \cite{18,19} comes from the two-axion
exchange.

Using the same approach, in \cite{20} stronger constraints on
axion-nucleon coupling constants over the wide range of axion
masses from $3\times 10^{-5}\,$eV to 1\,eV were obtained from
measurements of the Casimir force gradient between a sphere and
a plate coated with nonmagnetic and magnetic metals performed
by means of dynamic atomic force microscope \cite{21,22,23,24,25}.
 The strengthening up to a factor of 170, as compared to the
constraints of \cite{19}, was achieved. This demonstrates that
various experiments on measuring the Casimir interaction
\cite{26} are promising for further constraining the parameters
of an axion. In the past, these experiments were successfully
used to obtain stronger constraints on the Yukawa-type
corrections to Newtonian gravity due to exchange of light
scalar particles \cite{27} and from extra-dimensional physics
with low-energy compactification scale \cite{28} (see review
\cite{29} and the most recent results \cite{30,31,32,33,34,35}).

In this paper, we obtain stronger constraints on the pseudoscalar
coupling constants of axion-like particles to a proton and a
neutron
from measurements of the effective Casimir pressure by means of
micromechanical torsional oscillator \cite{36,37}. For this
purpose, we calculate the additional effective pressure in the
configuration of two parallel plates arising due to two-axion
exchange between a sphere and a plate (note that the experimental
 configuration \cite{36,37} involves a sphere oscillating in
the perpendicular direction to the plate, so that the effective
 pressure arises in the proximity force approximation
 \cite{26,29}).
The stronger limits on axion-nucleon coupling constants are
obtained over the range of axion masses from $10^{-3}\,$eV
to 15\,eV. The strengthening by factors from 2.2 to 3.2
in comparison with the limits of \cite{20} is achieved over
the range of axion masses from $10^{-3}\,$eV to 1\,eV,
respectively.
As compared to the limits of \cite{19}, the obtained constraints
are stronger up to a factor of 380.
Our model-independent constraints are
applicable on equal terms to axions and axion-like
particles. Because of this, below both terms are used
synonymously. All equations are written in the system of
units with $\hbar=c=1$.

\section{Pressure between two metallic plates due to two-axion
exchange}

In the experiment \cite{36,37}, the effective Casimir pressure
between two Au plates was determined from dynamic measurements
using a micromechanical torsional oscillator. The oscillator
consisted of a heavily doped polysilicon plate of area
$500\times 500\,\mu\mbox{m}^2$ and thickness $D=5\,\mu$m
suspended at two opposite points above the platform at the height
of about $2\,\mu$m. Two independent electrodes located on the
platform under the plate were used to measure the capacitance
between the electrodes and the plate. They were also used to
induce oscillation in the plate at the resonance frequency
of the micromachined oscillator. A large sapphire sphere
coated with layers of Cr and Au was attached to the
optical fiber above the oscillator. The sphere radius was
measured to be $R=151.3\,\mu$m. A silicon plate below the
sphere was also coated with layers of Cr and Au.

In the dynamic measurements, the vertical separation between
the sphere and the plate was varied harmonically with the
resonance frequency of oscillator, $\omega_r$, in the presence
of the sphere. The Casimir force between the sphere and the
plate caused the difference between $\omega_r$ and the
natural frequency of the oscillator $\omega_0$.
This difference has been measured and recalculated into
the gradient of the Casimir force acting
between the sphere and the plate,
$F_{sp}^{\prime}(a)$, using the solution for the linear
oscillator motion ($a$ is the absolute sphere-plate
separation). According to the proximity force approximation
(PFA) \cite{26,29},
\begin{equation}
F_{sp}(a)=2\pi RE(a),
\label{eq1}
\end{equation}
\noindent
where $E(a)$ is the Casimir energy per unit area of two
parallel plates (semispaces). Calculating the negative
derivative of both sides of (\ref{eq1}), one obtains the
effective Casimir pressure between two parallel plates
\begin{equation}
P(a)=-\frac{1}{2\pi R} F_{sp}^{\prime}(a),
\label{eq2}
\end{equation}
\noindent
which is the physical quantity indirectly measured
in \cite{36,37}. Note that under the condition $a\ll R$
the relative error in the gradient of the Casimir force
computed using (\ref{eq1}) does not exceed
$(0.3-0.4)a/R$ \cite{38,38a,39,40,41}.
Taking into account that below we consider separations
$a<300\,$nm, this is of less than 0.1\% error.

Now we calculate the additional pressure between two
parallel semispaces separated with a gap $a$ due to
two-axion exchange between nucleons. In this section, we
consider homogeneous semispaces and postpone the account
of finite thickness of the plate and layer structure of
both test bodies to Secs.~3 and 4.
First we perform a direct derivation of the additional
pressure $P_{\rm add}(a)$ by summing up the energies of
pair nucleon-nucleon interactions over the two semispaces
and calculating the negative derivative of the obtained
result. This pressure
can be considered as an addition to the
indirectly measured Casimir pressure (\ref{eq2}) if the
additional force between a sphere and a plate due to
two-axion exchange is related to the additional energy
per unit area of two parallel plates by the PFA,
so that
\begin{eqnarray}
&&
F_{sp,{\rm add}}(a)=2\pi RE_{\rm add}(a),
\label{eq3}\\
&&
P_{\rm add}(a)=-\frac{1}{2\pi R} F_{sp,{\rm add}}^{\prime}(a).
\nonumber
\end{eqnarray}
\noindent
Then we determine the application region of (\ref{eq3}) from
the comparison with the exact result for
$F_{sp,{\rm add}}^{\prime}(a)$.

Let the coordinate plane $x,\,y$ coincide with the boundary plane
of the lower semispace and let the $z$ axis be perpendicular to
it. The effective potential due to two-axion exchange between
two nucleons (protons or neutrons) situated at the points
$\mbox{\boldmath$r$}_1$ and $\mbox{\boldmath$r$}_2$ of the
upper and lower semispaces, respectively, is given by
\cite{18,42,43}
\begin{equation}
V_{kl}(|\mbox{\boldmath$r$}_1-\mbox{\boldmath$r$}_2|)=
-\frac{g_{ak}^2g_{al}^2\,m_a}{32\pi^3m^2}\,
\frac{K_1(2m_a|\mbox{\boldmath$r$}_1-
\mbox{\boldmath$r$}_2|)}{(\mbox{\boldmath$r$}_1-\mbox{\boldmath$r$}_2)^2}
.
\label{eq4}
\end{equation}
\noindent
Here, $g_{ak}$ and $g_{al}$ are the coupling constants of
an axion to a proton ($k,\,l=p$) or a neutron ($k,\,l=n$)
interaction, $m=(m_n+m_p)/2$ is the mean of the neutron and
proton masses, and $K_1(z)$ is the modified Bessel function
of the second kind. Equation (\ref{eq4}) was
derived under the condition
$|\mbox{\boldmath$r$}_1-\mbox{\boldmath$r$}_2|\gg 1/m$.
Taking into acount that in the experiment \cite{36,37}
we have $a>160\,$nm,
this condition is satisfied with large safety margin.

The additional energy per unit area of
the two semispaces due to
two-axion exchange can be written as
\begin{eqnarray}
&&
E_{\rm add}(a)=2\pi\sum_{k,l}n_{k,1}n_{l,2}
\int_{a}^{\infty}\!\!\!dz_1
\int_{-\infty}^{0}\!\!\!dz_2
\int_{0}^{\infty}\!\!\!\rho d\rho
\nonumber \\
&&~~~~~~~~~~~~~~~~~~~~
\times
V_{kl}(\sqrt{\rho^2+(z_1-z_2)^2}),
\label{eq5}
\end{eqnarray}
\noindent
where $V_{kl}$ is defined in (\ref{eq4}) and
\begin{equation}
n_{p,i}=\frac{\rho_i}{m_{\rm H}}\,\frac{Z_i}{\mu_i},
\qquad
n_{n,i}=\frac{\rho_i}{m_{\rm H}}\,\frac{N_i}{\mu_i}.
\label{eq5a}
\end{equation}
\noindent
Here $i=1,\,2$ numerates semispaces,
$\rho_{1,2}$ are the respective densities, $Z_{1,2}$
and $N_{1,2}$ are the numbers of protons and the mean number of
neutrons in the atoms (molecules) of
respective semispaces. The quantities
$\mu_{1,2}$ are given by
$\mu_{1,2}=m_{1,2}/m_{\rm H}$, where $m_{1,2}$ and $m_{\rm H}$
are the mean masses of the atoms (molecules) of the semispaces and
the mass of the atomic hydrogen, respectively.
The values of $Z/\mu$ and $N/\mu$ for the first 92 elements of
the Periodic Table with account of their isotopic composition
can be found in \cite{27}.

Calculating the negative derivative of (\ref{eq5}) with
respect to $a$, one obtains the additional pressure
between two semispaces
\begin{equation}
P_{\rm add}(a)=-\frac{m_a}{m^2m_{\rm H}^2}C_1C_2
\frac{\partial}{\partial a}
\int_{a}^{\infty}\!\!\!dz_1\,I(z_1),
\label{eq6}
\end{equation}
\noindent
where
\begin{equation}
I(z_1)\equiv\int_{-\infty}^{0}\!\!\!dz_2
\int_{0}^{\infty}\!\!\!\rho d\rho
\frac{{ K}_1(2m_a\sqrt{\rho^2+(z_1-z_2)^2})}{\rho^2+(z_1-z_2)^2}.
\label{eq7}
\end{equation}
\noindent
Here, the coefficients $C_{1,2}$ for the materials of the semispaces
are defined as
\begin{equation}
C_{1,2}=\rho_{1,2}\left(\frac{g_{ap}^2}{4\pi}\,
\frac{Z_{1,2}}{\mu_{1,2}}+\frac{g_{an}^2}{4\pi}\,
\frac{N_{1,2}}{\mu_{1,2}}\right).
\label{eq8}
\end{equation}
\noindent

 Using the integral representation
\cite{44}
\begin{equation}
\frac{K_1(z)}{z}=\int_{1}^{\infty}\!\!\!du\sqrt{u^2-1}e^{-zu}
\label{eq9}
\end{equation}
\noindent
and introducing the new variable $v=\sqrt{\rho^2+(z_1-z_2)^2}$,
one can rearrange (\ref{eq7}) into the form
\begin{equation}
I(z_1)=2m_a\int_{-\infty}^{0}\!\!\!dz_2
\int_{z_1-z_2}^{\infty}\!\!\!dv
\int_{1}^{\infty}\!\!\!du\sqrt{u^2-1}e^{-2m_auv}.
\label{eq10}
\end{equation}
\noindent
By integrating here with respect to $v$ and $z_2$, we arrive at
\begin{equation}
I(z_1)=\frac{1}{2m_a}
\int_{1}^{\infty}\!\!\!du\frac{\sqrt{u^2-1}}{u^2}e^{-2m_az_1u}.
\label{eq11}
\end{equation}
\noindent
Substituting this in (\ref{eq6}) and differentiating with
respect to $a$, we finally obtain
\begin{eqnarray}
P_{\rm add}(a)&=&
-\frac{1}{2\pi R}F_{sp,{\rm add}}^{\prime}(a)
\nonumber \\
&=&-\frac{C_1C_2}{2m^2m_{\rm H}^2}
\int_{1}^{\infty}\!\!\!du\frac{\sqrt{u^2-1}}{u^2}e^{-2m_aau}.
\label{eq12}
\end{eqnarray}

Now we determine the application region of (\ref{eq12}) in the
experimental configuration of \cite{36,37} which involves not
the two parallel plates, but a sphere above a plate.
By summing the potential (\ref{eq4}) over the volumes of a sphere
and a semispace it was shown \cite{20} that
\begin{eqnarray}
&&
-\frac{1}{2\pi R}F_{sp,{\rm add}}^{\prime}(a)=-\frac{C_sC_p}{2Rm^2m_{\rm H}^2}
\int_{1}^{\infty}\!\!\!du\frac{\sqrt{u^2-1}}{u^2}
\nonumber \\
&&~~~~~~~~~~~~~~~~~
\times
e^{-2m_aau}
\Phi(R,m_au),
\label{eq13}
\end{eqnarray}
\noindent
where the function $\Phi(r,z)$ is defined as
\begin{equation}
\Phi(r,z)=r-\frac{1}{2z}+e^{-2rz}\left(
r+\frac{1}{2z}\right)
\label{eq14}
\end{equation}
\noindent
and $C_s$ and $C_p$ are the constants for the sphere and plate
materials as defined in (\ref{eq8}). {}From (\ref{eq14})
we can see that
\begin{eqnarray}
&&
~~~~~~~~\frac{\Phi(R,m_au)}{R}=1-\frac{1}{2Rm_au}
\nonumber \\
&&~~~~~~~~~~~~~~~~~~
+e^{-2Rm_au}\left(
1+\frac{1}{2Rm_au}\right).
\label{eq15}
\end{eqnarray}
\noindent
Thus, (\ref{eq13}) leads to approximately the same results as
(\ref{eq12}) under the condition $Rm_a\gg 1$. Numerical
computations show that (\ref{eq12}) and (\ref{eq13}) deviate
less than approximately 1\% under the condition  $Rm_a> 10$.
Because of this, for the experimental parameters of \cite{36,37},
(\ref{eq12}) can be used for axion masses $m_a>10^{-2}\,$eV.
For smaller masses, calculations of forces due to two-axion
exchange using the PFA become not sufficiently exact.
In this case one should compute the additional effective pressure
using (\ref{eq13}). Note that similar results concerning the
application region of the PFA to Yukawa-type forces are obtained
in \cite{45,46}.

\section{Estimation of boundary effects}

Here, we consider the sphere above the plate of finite thicknes
and finite area and estimate errors in the additional force
gradient arising from treating this plate as infinitely large.
For convenience in calculations, we replace the square of
the area
$500\times 500\,\mu\mbox{m}^2$ by the disc of radius
$L=250\,\mu$m. The replacement of a square by a disc of smaller
area may only increase the boundary effects which, as we show
below, are sufficiently small.

By summing the potential (\ref{eq4}) over the volumes of a sphere
and a plate (disc) of thickness $D$ and radius $L$, an
additional contribution
due to the two-axion exchange to the quantity measured in
\cite{36,37} can be presented in the form \cite{20}
\begin{eqnarray}
&&
-\frac{1}{2\pi R}F_{\rm add}^{\prime}(a)=
-\frac{m_aC_sC_p}{2Rm^2m_{\rm H}^2}
\label{eq16} \\
&&~~
\times
\frac{\partial}{\partial a}
\int_{a}^{2R+a}\!\!\!dz_1\left[R^2-(R+a-z_1)^2\right]
G(z_1,m_a),
\nonumber
\end{eqnarray}
\noindent
where
\begin{eqnarray}
&&
G(z_1,m_a)\equiv\frac{\partial}{\partial z_1}
\int_{-D}^{0}\!\!\!\!dz_2
\int_{0}^{L}\!\!\!\!\rho d\rho
\nonumber \\
&&~~~~~~~~~~~~~~~~
\times\frac{{K}_1(2m_a\sqrt{\rho^2+(z_1-z_2)^2})}{\rho^2+(z_1-z_2)^2}.
\label{eq17}
\end{eqnarray}
\noindent
Using (\ref{eq9}), introducing the variable $v$ defined above and
integrating with respect to it, we obtain
\begin{eqnarray}
&&
G(z_1,m_a)=\int_{1}^{\infty}\!\!du\frac{\sqrt{u^2-1}}{u}
\frac{\partial}{\partial z_1}\int_{-D}^{0}\!\!\!dz_2
\nonumber \\
&&~~
\times\left[e^{-2m_au(z_1-z_2)}\right.
\left.-e^{-2m_au\sqrt{L^2+(z_1-z_2)^2}}\,\right].
\label{eq18}
\end{eqnarray}
\noindent
After integrating and differentiating in (\ref{eq18}) over $z_2$ and
$z_1$, respectively, we get
\begin{eqnarray}
&&
G(z_1,m_a)=\int_{1}^{\infty}\!\!du\frac{\sqrt{u^2-1}}{u}
\left[e^{-2m_aau}\left(1-e^{-2m_aDu}\right)\right.
\nonumber \\
&&~~~~~~~~
\left.-e^{-2m_au\sqrt{L^2+z_1^2}}+
e^{-2m_au\sqrt{L^2+(z_1+D)^2}}\,\right].
\label{eq19}
\end{eqnarray}

Now we substitute (\ref{eq19}) in  (\ref{eq16}).
In doing so, we integrate only the first term on the right-hand
side of (\ref{eq19}) with respect to $z_1$ and perform the
differentiation with respect to $a$. The result is
\begin{eqnarray}
&&
-\frac{1}{2\pi R}F_{\rm add}^{\prime}(a)=
-\frac{C_sC_p}{2Rm^2m_{\rm H}^2}
\int_{1}^{\infty}\!\!du\frac{\sqrt{u^2-1}}{u^2}
\nonumber \\
&&~
\times\left[e^{-2m_aau}
\left(1-e^{-2m_aDu}\right)\Phi(R,m_au)
\right.
\nonumber \\
&&~~~~~~~~~~~~~~~~~~
\left.\vphantom{\left(e^{-2m_aDu}\right)}
-
Y(m_au,L,D)\right],
\label{eq20}
\end{eqnarray}
\noindent
where the function $\Phi(r,z)$ is defined in (\ref{eq15}) and
the following notation is introduced
\begin{eqnarray}
&&
Y(m_au,L,D)\equiv 2m_au\int_{a}^{2R+a}\!\!dz_1
(R+a-z_1)
\nonumber \\
&&~~
\times
\left[e^{-2m_au\sqrt{L^2+z_1^2}}-
e^{-2m_au\sqrt{L^2+(z_1+D)^2}}\,
\right].
\label{eq21}
\end{eqnarray}

{}From the comparison of the right-hand sides of (\ref{eq20})
and (\ref{eq13}), it is seen that the first term of
(\ref{eq20}) generalizes (\ref{eq13}) for the case of a
plate of finite thickness $D$. In the limiting case
$D\to\infty$ the first term of (\ref{eq20}) coincides with
(\ref{eq13}). The second term on the right-hand side of
(\ref{eq20}) takes into account the boundary effects.

Now we estimate the relative role of boundary effects in the
calculation of the additional force gradient due to two-axion
exchange using the experimental parameters of \cite{36,37}.
Taking into account that the quantity in square brackets
on the right-hand side of (\ref{eq21}) is positive, one can
only increase the integral by omitting the part of the
integration domain
where the quantity in the round brackets is
negative. This results in the inequality
\begin{eqnarray}
&&
Y(m_au,L,D)< 2m_au\int_{a}^{R+a}\!\!dz_1
(R+a-z_1)
\nonumber \\
&&~~
\times
\left[e^{-2m_au\sqrt{L^2+z_1^2}}-
e^{-2m_au\sqrt{L^2+(z_1+D)^2}}\,
\right].
\label{eq22}
\end{eqnarray}
\noindent
The second exponent on the right-hand side of this equation
under the condition $D\ll L$ can be approximated as
\begin{eqnarray}
&&
e^{-2m_au\sqrt{L^2+(z_1+D)^2}}
\approx
e^{-2m_au\sqrt{L^2+z_1^2}}\,
e^{-m_au\frac{2z_1D+D^2}{\sqrt{L^2+z_1^2}}}
\nonumber \\
&&~~~~~~
\approx
e^{-2m_au\sqrt{L^2+z_1^2}}\left(1-
m_au\frac{2z_1D+D^2}{\sqrt{L^2+z_1^2}}\right),
\label{eq23}
\end{eqnarray}
\noindent
where the last transformation is performed for small axion
masses $m_a\sim 1/R$ leading to the largest boundary effects
[the dominant contribution to the integral (\ref{eq20}) is
given by $u\sim 1$]. Substituting (\ref{eq23}) in (\ref{eq22}),
one obtains
\begin{eqnarray}
&&
Y(m_au,L,D)< 2(m_au)^2\int_{a}^{R+a}\!\!dz_1
(R+a-z_1)
\nonumber \\
&&~~~~~
\times
e^{-2m_au\sqrt{L^2+z_1^2}}
\frac{2z_1D+D^2}{\sqrt{L^2+z_1^2}}
\nonumber \\
&&~~
<2(m_au)^2e^{-2m_auL}\frac{D}{L}
\nonumber \\
&&~~~~~~~~~~~~~~~
\times
\int_{a}^{R+a}\!\!\!dz_1
(R+a-z_1)(2z_1+D)
\nonumber \\
&&~~
\approx\frac{2RD}{3L}(m_aRu)^2e^{-2m_auL}.
\label{eq24}
\end{eqnarray}

{}From (\ref{eq24}) it is seen that
under the conditions
$m_aR\approx 1$ and $u\sim 1$ it follows:
\begin{equation}
Y(m_au,L,D)<3\times 10^{-4}R.
\label{eq25}
\end{equation}

On the same conditions, the contribution of the remaining terms
in the square brackets of (\ref{eq20}) is equal to
$3\times 10^{-2}R$. Thus, the boundary effects contribute less
than 1\% under the integral (\ref{eq20}). We have checked by means
of numerical computations that the contribution of the boundary
effects to the normalized gradient of the additional force also
does not exceed 1\%. Because of this, the role of additional
forces due to two-axion exchange in the experiment \cite{36,37}
can be calculated under the assumption of the infinitely large
area of the oscillator plate.

\section{Account of layer structure of test bodies}

As was mentioned in Sec.~2, the test bodies in the experiment
\cite{36,37} were not homogeneous. The Si plate of an oscillator
of finite thickness $D$ was coated with a Cr layer of thickness
$\Delta_p^{\!\rm Cr}=10\,$nm and with an outer Au layer
of thickness $\Delta_p^{\!\rm Au}=210\,$nm. The sapphire
(Al${}_2$O${}_3$) sphere was coated with a Cr layer of thickness
$\Delta_s^{\!\rm Cr}=10\,$nm and then with an Au layer
of thickness $\Delta_s^{\!\rm Au}=180\,$nm.
The densities of all these materials are presented in the second
column of Table~1.
\begin{table}[h]
\caption{The values of densities (column 2) and quantities
$Z/\mu$ (column 3) and $N/\mu$ (column 4) are presented for
different materials (column 1). See text for further
discussion.
}
\label{tab:1}       
\begin{center}
\begin{tabular}{cccc}
\hline\noalign{\smallskip}
Material&$\rho\,(\mbox{g/cm}^3)$&$\frac{Z}{\mu}$&
$\frac{N}{\mu}$ \\
\noalign{\smallskip}\hline\noalign{\smallskip}
Au &19.28 & 0.40422 & 0.60378 \\
Cr& 7.15 & 0.46518 & 0.54379 \\
Si & 2.33 & 0.50238 & 0.50628 \\
Al${}_2$O${}_3$ & 4.1 & 0.49422 & 0.51412 \\
\noalign{\smallskip}\hline
\end{tabular}
\end{center}
\end{table}

Now we adapt the results of Sec.~2 for the additional effective
pressure due to two-axion exchange for the case of experimental
layer structure of both bodies and finite thickness of the
oscillator plate. We begin with (\ref{eq12}), which can be used
in the experimental configuration of \cite{36,37} within the
application region of the PFA. The layers are taken into account
one by one. For instance, to account for the Au layer on the
plate, we subtract from (\ref{eq12}), written for two Au
semispaces, the effective pressure between the same semispaces,
but separated by the gap $a+\Delta_p^{\!\rm Au}$. Then we add
the effective pressure for Au-Cr semispaces separated by the
same gap and subtract the pressure for these semispaces
separated by the gap $a+\Delta_p^{\!\rm Au}+\Delta_p^{\!\rm Cr}$
etc. Similar procedure is used to account for the layer
structure of the upper plate. Finally, for the experimental
configuration one obtains
\begin{eqnarray}
&&
P_{\rm add}(a)=
-\frac{1}{2\pi R}F_{sp,{\rm add}}^{\prime}
=
-\frac{1}{2m^2m_{\rm H}^2}\int_{1}^{\infty}\!\!\!du
\frac{\sqrt{u^2-1}}{u^2}
\nonumber \\
&&~~~~~~~~~~
\times
e^{-2m_aau}X_p(m_au)X_s(m_au),
\label{eq26}
\end{eqnarray}
\noindent
where
\begin{eqnarray}
&&
X_p(m_au)\equiv C_{\rm Au}\left(1-e^{-2m_au\Delta_p^{\!\rm Au}}
\right)
\nonumber \\
&&~~~
+C_{\rm Cr}e^{-2m_au\Delta_p^{\!\rm Au}}
\left(1-e^{-2m_au\Delta_p^{\!\rm Cr}}
\right)
\nonumber \\
&&~~~
+C_{\rm Si}e^{-2m_au(\Delta_p^{\!\rm Au}+\Delta_p^{\!\rm Cr})}
\left(1-e^{-2m_auD}
\right),
\nonumber \\[1mm]
&&
X_s(m_au)\equiv C_{\rm Au}\left(1-e^{-2m_au\Delta_s^{\!\rm Au}}
\right)
\nonumber \\
&&~~~
+C_{\rm Cr}e^{-2m_au\Delta_s^{\!\rm Au}}
\left(1-e^{-2m_au\Delta_s^{\!\rm Cr}}
\right)
\nonumber \\
&&~~~
+C_{\rm Al_2O_3}
e^{-2m_au(\Delta_s^{\!\rm Au}+\Delta_s^{\!\rm Cr})}.
\label{eq27}
\end{eqnarray}
\noindent
Here, the coefficients $C_{\rm Au}$,  $C_{\rm Cr}$
and $C_{\rm Si}$ are defined in (\ref{eq8}).
They are calculated using the respective values for
$Z/\mu$ and $N/\mu$ presented in the third and fourth
columns of Table~1 \cite{27}. The quantities
$Z/\mu$ and $N/\mu$ for Al${}_2$O${}_3$ are also
given in Table~1 \cite{20}.

As was found in Sec.~2, in the experimental configuration
\cite{36,37}, the PFA is applicable to calculate additional
forces due to two-axion exchange under the condition
$m_a>10^{-2}\,$eV. For axions of smaller masses a more
exact expression (\ref{eq13}) should be used. It can be
adapted for the experimental layer structure using the
procedure described above. The result is
\begin{eqnarray}
&&
-\frac{1}{2\pi R}F_{sp,{\rm add}}^{\prime}=
-\frac{1}{2m^2m_{\rm H}^2R}\int_{1}^{\infty}\!\!\!du
\frac{\sqrt{u^2-1}}{u^2}
\nonumber \\
&&~~~~~~~~~
\times
e^{-2m_aau}X_p(m_au)\tilde{X}_s(m_au),
\label{eq28}
\end{eqnarray}
\noindent
where the function $\tilde{X}_s$ is defined as
\begin{eqnarray}
&&
\tilde{X}_s(m_au)\equiv C_{\rm Au}\left[
\vphantom{e^{-2m_au\Delta_s^{\!\rm Au}}}
\Phi(R,m_au)
\right.
\label{eq29}\\
&&~~~~~~~~~~~~
\left.
-e^{-2m_au\Delta_s^{\!\rm Au}}
\Phi(R-\Delta_s^{\!\rm Au},m_au)\right]
\nonumber \\
&&~~~
+C_{\rm Cr}e^{-2m_au\Delta_s^{\!\rm Au}}
\left[
\vphantom{e^{-2m_au\Delta_s^{\!\rm Au}}}
\Phi(R-\Delta_s^{\!\rm Au},m_au)
\right.
\nonumber \\
&&~~~~~~~~~~~~
\left.
-e^{-2m_au\Delta_s^{\!\rm Cr}}
\Phi(R-\Delta_s^{\!\rm Au}-\Delta_s^{\!\rm Cr},m_au)
\right]
\nonumber \\
&&~~~
+C_{\rm Al_2O_3}
e^{-2m_au(\Delta_s^{\!\rm Au}+\Delta_s^{\!\rm Cr})}
\Phi(R-\Delta_s^{\!\rm Au}-\Delta_s^{\!\rm Cr},m_au).
\nonumber
\end{eqnarray}
\noindent
Here, the functions $X_p$ and $\Phi$ are given in (\ref{eq27})
and (\ref{eq14}), respectively.

\section{Constraints on axion-nucleon coupling constants}

The experimental data of \cite{36,37} for the effective
Casimir pressure were obtained at separations $a>160\,$nm
and found to be in good agreement with the Lifshitz theory
\cite{47} under the condition that the low-frequency
behavior of the dielectric permittivity of Au is described
by the plasma model (the Casimir force is entirely determined by
the outer Au layers on both test bodies and, as opposed to the
additional force due to two-axion exchange, is not influenced
by the layers situated below). No signature of any additional
interaction was observed in the limits of the total
experimental error, $\Delta P(a)$, in the pressure measurements.

This means that the effective additional pressure should satisfy
the following inequality:
\begin{equation}
\left|-\frac{1}{2\pi R}F_{sp,{\rm add}}^{\prime}\right|
\leq\Delta P(a).
\label{eq30}
\end{equation}
\noindent
The left-hand side of this inequality is given by the magnitudes
of either (\ref{eq26}) (for axion masses allowing the use of the
PFA) or (\ref{eq28}) (for axion of smaller masses).
The total experimental error in the indirectly measured
pressures, $\Delta P(a)$, recalculated with the 67\% confidence
level for convenience in comparison with the previously obtained
constraints, is equal to 0.55, 0.38, and 0.22\,mPa at
separations $a=162$, 200, and 300\,nm, respectively.

\begin{figure}[t]
\vspace*{-5.6cm}
\resizebox{0.6\textwidth}{!}{%
\hspace*{-3.3cm} \includegraphics{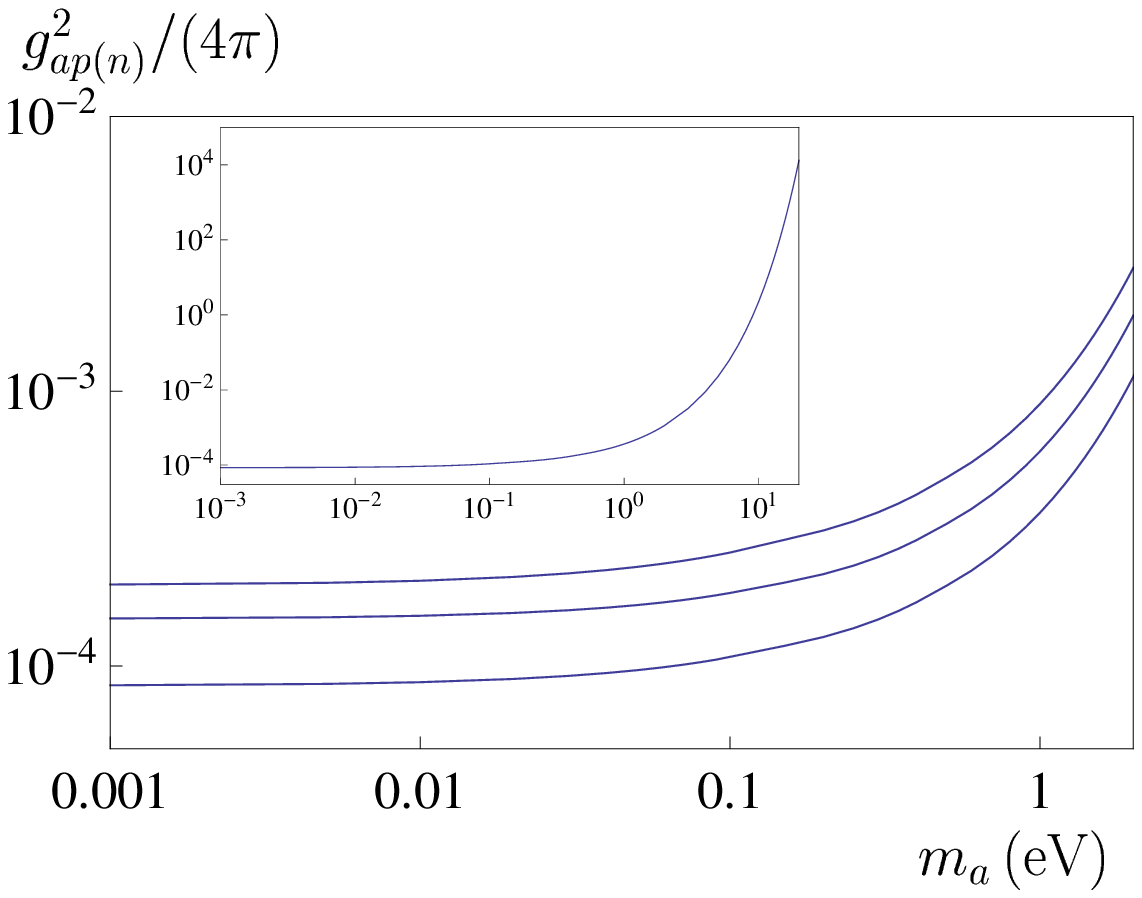}
}
\vspace*{-6cm}
\caption{Constraints on the coupling constants of an axion to
a proton or a neutron obtained from indirect measurements of
the effective Casimir pressure versus the axion mass.
The lines from bottom to top are plotted
under the conditions
$g_{ap}^2=g_{an}^2$, $g_{an}^2\gg g_{ap}^2$, and
$g_{ap}^2\gg g_{an}^2$, respectively.
In an inset the line is plotted under the condition
$g_{ap}^2=g_{an}^2$ for larger masses.
The regions of the plane above each line are prohibited and
below each line are allowed.}
\label{fig:1}       
\end{figure}
We have found numerically (see Fig.~1)
the values of the axion to nucleon
coupling constants $g_{ap},\,g_{an}$ and masses $m_a$
satisfying the inequality (\ref{eq30}). For this purpose, the
expressions (\ref{eq26}) and (\ref{eq28}) were substituted in
(\ref{eq30}) over the mass intervals
$10^{-2}\,\mbox{eV}<m_a<15\,$eV and
$10^{-3}\,\mbox{eV}<m_a<10^{-2}\,$eV, respectively.
We do not consider the axion masses $m_a<10^{-3}\,$eV
because in this case the respective Compton wavelengths become
too large and one cannot neglect the role of boundary
effects (see Sec.~3). For $m_a>15\,$eV the constraints on
$g_{ap}$ and $g_{an}$ following from this experiment become
much weaker. In different intervals of $m_a$, the strongest
constraints follow from the inequality (\ref{eq30}) considered
at different separation distances. Thus, for $m_a<0.1\,$eV
the strongest constraints result at $a=300\,$nm and for
$0.1\,\mbox{eV}\leq m_a<0.5\,$eV and
$0.5\,\mbox{eV}\leq m_a<15\,$eV at $a=200\,$nm and
162\,nm, respectively.

In Fig.~1, we present the obtained strongest constraints on the
constants $g_{ap(n)}^2/(4\pi)$ as functions of the axion mass
$m_a$.
The lines correspond to the equality sign in (\ref{eq30}).
In Fig.~1 the three lines from bottom to top are plotted
under the conditions
$g_{ap}^2=g_{an}^2$, $g_{an}^2\gg g_{ap}^2$, and
$g_{ap}^2\gg g_{an}^2$, respectively, for axion masses below
2\,eV.
The regions of the $(m_a,g_{ap(n)}^2)$ plane above each line
are prohibited by the results of experiment \cite{36,37},
because the coordinates of their points violate inequality
(\ref{eq30}). The regions below each line are allowed by the
results of this experiment. As can be seen in Fig.~1,
for axions with masses $m_a<10^{-2}\,$eV the obtained constraints
are almost independent of $m_a$.
In an inset to Fig.~1 we plot the obtained constraints over a wider range
of $m_a$ (up to 15\,eV) under the condition $g_{ap}^2=g_{an}^2$.
As is seen in this figure, with increasing $m_a$ the strength
of constraints quickly decreases.
In Table~2, we present the maximum allowed values
of the axion-nucleon
coupling constants over the most interesting region of masses
from $m_a=10^{-3}\,$eV to 2\,eV (column 1) partially overlapping
with an axion window. The values in column 2 are obtained under
 the conditions $g_{ap}^2=g_{an}^2$, and columns 3 and 4 contain
 the maximum values of $g_{an}^2/(4\pi)$ and $g_{ap}^2/(4\pi)$
found under the conditions
 $g_{an}^2\gg g_{ap}^2$ and $g_{ap}^2\gg g_{an}^2$,
 respectively.

\begin{table}[b]
\caption{Maximum values of the coupling constants of
an axion to a proton and a neutron, allowed by
indirect measurements of the Casimir pressure
between Au plates, are calculated for different axion
masses (column 1) under the conditions
$g_{ap}^2=g_{an}^2$ (column 2),
$g_{an}^2\gg g_{ap}^2$ (column 3),
and $g_{ap}^2\gg g_{an}^2$ (column 4).
}
\label{tab:2}       
\begin{center}
\begin{tabular}{cccc}
\hline\noalign{\smallskip}
$m_a\,$(eV) & $\frac{g_{ap}^2}{4\pi}=\frac{g_{an}^2}{4\pi}$ &
$\frac{g_{an}^2}{4\pi}\gg\frac{g_{ap}^2}{4\pi}$ &
$\frac{g_{ap}^2}{4\pi}\gg\frac{g_{an}^2}{4\pi}$ \\
\noalign{\smallskip}\hline\noalign{\smallskip}
$0.001$& $8.51\times 10^{-5}$&$1.49\times 10^{-4}$&
$1.98\times 10^{-4}$ \\
0.01 &$8.74\times 10^{-5}$&$1.52\times 10^{-4}$&
$2.04\times 10^{-4}$ \\
0.05 &$9.66\times 10^{-5}$&$1.67\times 10^{-4}$&
$2.30\times 10^{-4}$ \\
0.1 &$1.08\times 10^{-4}$&$1.84\times 10^{-4}$&
$2.59\times 10^{-4}$ \\
0.2&$1.28\times 10^{-4}$&$2.16\times 10^{-4}$&
$3.11\times 10^{-4}$ \\
0.3&$1.48\times 10^{-4}$&$2.49\times 10^{-4}$&
$3.62\times 10^{-4}$ \\
0.4&$1.71\times 10^{-4}$&$2.88\times 10^{-4}$&
$4.21\times 10^{-4} $\\
0.5&$1.96\times 10^{-4}$&$3.29\times 10^{-4}$&
$4.85\times 10^{-4} $\\
0.6&$2.22\times 10^{-4}$&$3.73\times 10^{-4}$&
$5.50\times 10^{-4} $\\
0.7&$2.51\times 10^{-4}$&$4.21\times 10^{-4}$&
$6.24\times 10^{-4} $\\
0.8&$2.84\times 10^{-4}$&$4.76\times 10^{-4}$&
$7.06\times 10^{-4} $\\
0.9&$3.21\times 10^{-4}$&$5.37\times 10^{-4}$&
$7.97\times 10^{-4} $\\
1.0&$3.62\times 10^{-4}$&$6.05\times 10^{-4}$&
$8.99\times 10^{-4}$ \\
1.5&$6.48\times 10^{-4}$&$1.08\times 10^{-3}$&
$1.61\times 10^{-3}$ \\
2.0&$1.13\times 10^{-3}$&$1.88\times 10^{-3}$&
$2.81\times 10^{-3}$\\
\noalign{\smallskip}\hline
\end{tabular}
\end{center}
\end{table}
In Fig.~2(a) the constraints derived in this paper are
compared with those found previously \cite{19,20}
from measurements of  the thermal Casimir-Polder force
\cite{48} and from experiments on
measuring the gradient of the Ca\-si\-mir force between
Au surfaces \cite{21,22}.  For the sake of definiteness,
the comparison is made under the most reasonable
condition $g_{ap}^2=g_{an}^2$. The solid line in
Fig.~2(a)
reproduces the lower line in Fig.~1 obtained here.
The dashed lines 1 and 2 reproduce the constraints
obtained \cite{19,20} from measurements of  the Casimir-Polder
force and the gradient of the Casimir force over the
regions of axion masses $m_a\leq 0.3\,$eV and $m_a\leq 1\,$eV,
respectively. The regions of the plane above each line are
prohibited and
below each line are allowed by the results of the respective
experiment.
\begin{figure}[t]
\vspace*{-3cm}
\resizebox{0.8\textwidth}{!}{%
\hspace*{-4.6cm} \includegraphics{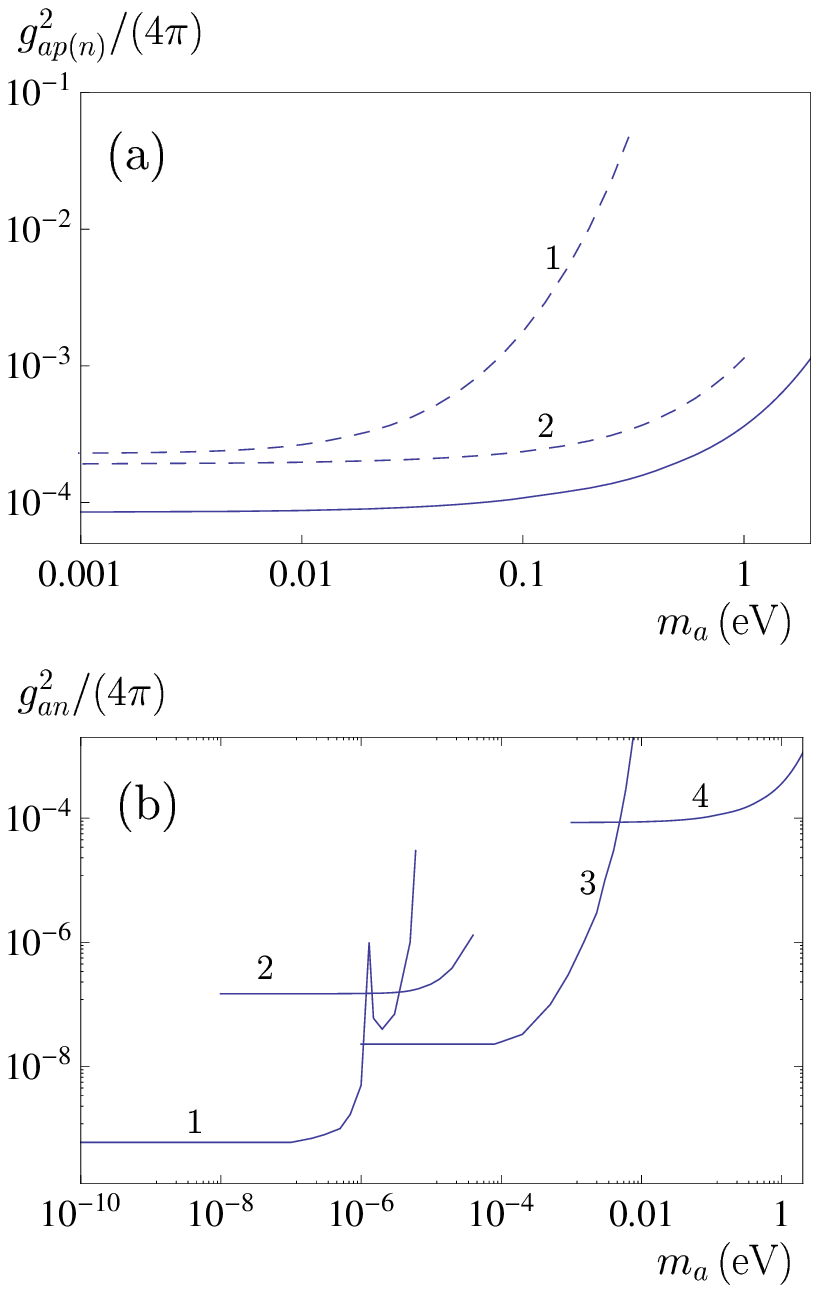}
}
\vspace*{-10.5cm}
\caption{(a) Comparison between the constraints on the coupling
constant of an axion to nucleon obtained here
under the condition $g_{ap}^2=g_{an}^2$ (the solid line)
with those obtained previously from experiments on
measuring the thermal Casimir-Polder force (the dashed line 1)
and the gradient of the Casimir force (the dashed line 2).
The regions of the plane above each line are prohibited and
below each line are allowed.
(b) Constraints on the coupling constant $g_{an}$ following
from magnetometer measurements \cite{53a} (the line 1), from
the Cavendish-type experiments \cite{17,21a} and \cite{53b}
(the lines 2 and 3, respectively), and obtained in this
work from measurements of the Casimir pressure by means of
micromachined oscillator \cite{36,37} (the line 4).
}
\label{fig:2}       
\end{figure}
As can be seen in Fig.~2(a), at $m_a=10^{-3}\,$eV and 1\,eV
our present constraints are stronger by the factors of 2.2
and 3.2, respectively, than those obtained from
measurements of the gradient of the Casimir force (the
dashed line 2). In comparison with the constraints from
measurements of the Casimir-Polder force (the dashed line 1),
the present constraints are stronger up to a factor of 380. This
strengthening is achieved for the axion mass $m_a=0.3\,$eV.

Now we compare the obtained here strongest model-independent
constraints on the coupling constant $g_{an}$
(the lower line in Fig.~1) with other model-independent
constraints obtained to the present day. The line 1 in Fig.~2(b)
shows the constraints found \cite{53a} with the help of a
magnetometer using spin-polarized K and ${}^3$He atoms.
These constraints are obtained in the region of axion masses
from $10^{-10}$ to $6\times 10^{-6}\,$eV.
The line 2 shows the constraints found in \cite{18} from the
Cavendish-type experiment \cite{17,21a} for $m_a$
from $10^{-9}$ to $10^{-5}\,$eV.
The results of a more modern Cavendish-type experiment \cite{53b}
were used to constrain $g_{an}$ in the region
from $m_a=10^{-6}\,$eV to $m_a=10^{-2}\,$eV \cite{53c}.
These results are shown by the line 3 in Fig.~2(b).
Our constraints obtained here are shown by the line 4.
As is seen in Fig.~2(b), the model-independent constraints
become weaker with increasing $m_a$ (the same takes place
for the constraints on Yukawa-type corrections to Newtonian
gravity arising from the exchange of scalar particles
\cite{29,30,31,32,33,34,35}). It can be seen, however, that
in the range of axion masses from $2\times 10^{-3}$ to 0.3\,eV
our constraints following from the Casimir effect are the
strongest model-independent constraints.

A lot of constraints on an axion were obtained using some model
approaches. Thus, the planar Si(Li) detector placed inside the
low-background setup was used to detect the $\gamma$-quanta
appearing in the deexcitation of the nuclear level excited
by a solar axion \cite{53d}. In the framework of the model of
hadronic axions, where the coupling constant is a function
of the mass, the upper limits for the axion mass
$m_a\leq 159\,$eV \cite{53d} and $m_a\leq 145\,$eV \cite{53e}
were obtained. {}From the neutrino data of supernova SN 1987A it
was found \cite{53f} that for the model of hadronic axions
$g_{ap(n)}<10^{-10}$ or $g_{ap(n)}>10^{-3}$ with a narrow
allowed region in the vicinity of $g_{ap(n)}=10^{-6}$.
{}From astrophysical arguments connected with stellar cooling
by the emission of hadronic axions a similar bound
$g_{ap(n)}<3\times 10^{-10}$ was obtained \cite{53g,53h}.
It should be noted, however, that the emission rate suffers from
significant uncertainties related to dense nuclear matter
effects \cite{53h}. In addition to a pseudoscalar coupling of
axions to nucleons, it is possible also to introduce the scalar
one \cite{53i} and consider respective coupling constants
$g_{ap(n)}^{(s)}$. Several constraints on the product of
constants $|g_{an}g_{an}^{(s)}|$ were obtained from
experiments on neutron diffraction \cite{53j}. Thus, it was
shown \cite{53j} that  $|g_{an}g_{an}^{(s)}|<10^{-11}$
within the range of axion masses
$2\times 10^{-5}\,\mbox{eV}<m_a<2\times 10^{3}\,$eV.
When the axion mass increases up to $2\times 10^6\,$eV,
the respective constraint becomes less stringent:
$|g_{an}g_{an}^{(s)}|<10^{-7}$.

At the end of this section, we note that subsequent independent
measurements of the gradient of the Casimir force in
\cite{21,22,23,24,25} confirmed both the experimental results
of \cite{36,37} and their agreement with the Lifshitz theory
under the condition that the low-frequency behavior of the
dielectric permittivity of Au is described by the plasma
model (the conclusion of \cite{49}, claiming an agreement
with the Drude model low-frequency behavior over the same range
of separations was shown \cite{50} to be based on an unaccounted
systematic error).

\section{Conclusions and discussion}

In this paper we have derived stronger constraints on the
pseudoscalar coupling constants of an axion to a proton and
a neutron from measurements of the effective Casimir pressure
by means of a micromachined oscillator. For this purpose,
we have calculated the additional pressure between two parallel
plates due to two-axion exchange and determined the validity
region of the PFA when it is applied to the forces of axion
origin. The role of boundary effects due to a finite area of
the oscillator plate was determined.

The obtained constraints are applicable over a wide region of
axion masses from $10^{-3}\,$eV to 15\,eV, partially overlapping
with an axion window. Under the assumption that $g_{ap}=g_{an}$,
they are stronger up to a factor of 380 than the previously
known laboratory constraints in this mass range derived from
measurements of the thermal Casimir-Polder force and up to a
factor of 3.15 than those found from measurements of the
gradient of the Casimir force by means of AFM.

The obtained results demonstrate that measurements of the
Casimir interaction using different laboratory techniques are
useful in searching axion-like particles and constraining
their coupling constants to nucleons. In future, it seems pro\-mi\-sing to
consider the potentialities of more complicated experimental
configurations, specifically, with corrugated boundary
surfaces, for obtaining stronger constraints on the parameters
of axion-like particles.

\begin{acknowledgement}
The authors of this work acknowledge CNPq (Brazil) for
 partial financial support.
G.L.K.\ and V.M.M.\ are grateful to M.\ Yu.\ Khlopov for
useful discussions. They also acknowledge
the Department
of Physics of the Federal University of
Para\'{\i}ba (Jo\~{a}o Pessoa, Brazil) for hospitality.
\end{acknowledgement}


\begin{thebibliography}{99}
\bibitem{1}
S.~Weinberg,
Phys. Rev. Lett. {\bf 40}, 223 (1978).
\bibitem{2}
F.~Wilczek,
Phys. Rev. Lett. {\bf 40}, 279 (1978).
\bibitem{3}
J.~E.~Kim, G.~Carosi,
Rev. Mod. Phys. {\bf 82}, 557 (2010).
\bibitem{4}
J.\ Beringer {\it et al.} (Particle Data Group),
Phys. Rev. D {\bf 86}, 010001 (2012).
\bibitem{5}
R.~D.~Peccei, H.~R.~Quinn,
Phys. Rev. Lett. {\bf 38}, 1440 (1977).
\bibitem{5a}
K.~Baker {\it et al.}
Ann. Phys. (Berlin) {\bf 525}, A93 (2013).
\bibitem{6}
J.~E.~Kim,
Phys. Rev. Lett. {\bf 43}, 103 (1979).
\bibitem{7}
M.~A.~Shifman, A.~I.~Vainstein, V.~I.~Zakharov,
Nucl. Phys. B {\bf 166}, 493 (1980).
\bibitem{8}
A.~P.~Zhitnitskii,
Sov. J. Nucl. Phys. {\bf 31}, 260 (1980).
\bibitem{9}
M.~Dine, F.~Fischler, M.~Srednicki,
Phys. Lett. B {\bf 104}, 199 (1981).
\bibitem{9a}
Z.~G.~Berezhiani, M.~Yu.~Khlopov,
Z.Phys. C --- Particles and Fields
{\bf 49}, 73 (1991).
\bibitem{9b}
M.~Khlopov,
{\it Fundamentals of Cosmic Particle Physics}
(CISP-Springer, Cambridge, 2012).
\bibitem{10}
A.~V.~Derbin, S.~V.~Bakhlanov, I.~S.~Dratchnev,
A.\ S.\ Kayunov, V.\ N.\ Muratova,
Eur. Phys. J. C {\bf 73}, 2490 (2013).
\bibitem{11}
J.~Jaeckel, E.~Mass\'{o}, J.~Redondo, A.~Ringwald,
F.~Takahashi,
Phys. Rev. D {\bf 75}, 013004 (2007).
\bibitem{12}
P.~Brax, C.~van~de~Bruck, A.-C.~Davis,
Phys. Rev. Lett. {\bf 99}, 121103 (2007).
\bibitem{12a}
J.~E.~Kim,
Phys. Rep. {\bf 150}, 1 (1987).
\bibitem{12b}
Yu.~N.~Gnedin,
Int. J. Mod. Phys. A {\bf 17}, 4251 (2002).
\bibitem{13}
E.~Fischbach, D.~E.~Krause,
{Phys. Rev. Lett.} {\bf 82}, 4753 (1999).
\bibitem{14}
E.~Fischbach, D.~E.~Krause,
{Phys. Rev. Lett.} {\bf 83}, 3593 (1999).
\bibitem{15}
G.~L.~Smith, C.~D.~Hoyle, J.~H.~Gundlach, E.~G.~Adelberger,
B.~R.~Heckel,  H.~E.~Swanson,
Phys. Rev. D {\bf 61}, 022001 (1999).
\bibitem{16}
J.~H.~Gundlach, G.~L.~Smith, E.~G.~Adelberger,
B.~R.~Heckel,  H.~E.~Swanson,
Phys. Rev. Lett. {\bf 78}, 2523 (1997).
\bibitem{17}
R.~Spero, J.~K.~Hoskins, R.~Newman, J.\ Pellam,
J.~Schultz,
Phys. Rev. Lett. {\bf 44}, 1645 (1980).
\bibitem{21a}
J.~K.~Hoskins, R.~D.~Newman, R.~Spero, J.\ Schulz,
Phys. Rev. D {\bf 32}, 3084 (1985).
\bibitem{18}
E.~G.~Adelberger, E.~Fischbach, D.~E.~Krause,  R.\ D.\ Newman,
{Phys. Rev. D} {\bf 68}, 062002 (2003).
\bibitem{19}
V.~B.~Bezerra, G.~L.~Klimchitskaya,
 V.~M.~Mostepanenko,  C.~Romero,
Phys. Rev. D {\bf 89}, 035010 (2014).
\bibitem{48}
J.~M.~Obrecht, R.~J.~Wild, M.~Antezza, L.~P.~Pitaevskii,
S.~Stringari, E.~A.~Cornell,
Phys. Rev. Lett. {\bf 98}, 063201 (2007).
\bibitem{20}
V.~B.~Bezerra, G.~L.~Klimchitskaya,
 V.~M.~Mostepanenko, C.~Romero,
arXiv:1402.2528; Phys. Rev. D, to appear.
\bibitem{21}
C.-C.~Chang, A.~A.~Banishev, R.~Castillo-Garza,
G.~L.~Klimchitskaya, V.\ M.\ Mostepanenko,  U.\ Mohideen,
Phys. Rev. B {\bf 85}, 165443 (2012).
\bibitem{22}
A.~A.~Banishev, C.-C.~Chang, R.~Castillo-Garza,
G.~L.~Klimchitskaya, V.\ M.\ Mostepanenko,  U.\ Mohideen,
Int. J. Mod. Phys. A {\bf 27}, 1260001 (2012).
\bibitem{23}
A.~A.~Banishev, C.-C.~Chang,
G.~L.~Klimchitskaya, V.\ M.\ Mostepanenko, U.\ Mohideen,
Phys. Rev. B {\bf 85}, 195422 (2012).
\bibitem{24}
A.~A.~Banishev,
G.~L.~Klimchitskaya, V.\ M.\ Mostepanenko, U.\ Mohideen,
Phys. Rev. Lett. {\bf 110}, 137401 (2013).
\bibitem{25}
A.~A.~Banishev,
G.~L.~Klimchitskaya, V.\ M.\ Mostepanenko,  U.\ Mohideen,
Phys. Rev. B {\bf 88}, 155410 (2013).
\bibitem {26}
G.~L.~Klimchitskaya, U. Mohideen, V.\ M.\ Mostepanenko,
Rev. Mod. Phys. {\bf 81}, 1827 (2009).
\bibitem{27}
E.~Fischbach, C.~L.~Talmadge, {\it The Search for Non-Newtonian
Gravity} (Springer, New York, 1999).
\bibitem{28}
I.~Antoniadis,
N.~Arkani-Hamed, S.~Dimopoulos, G.~Dvali,
Phys. Lett. B {\bf 436}, 257 (1998).
\bibitem{29}
M.~Bordag, G.~L.~Klimchitskaya, U.\ Mohideen,
V.\ M.\ Mostepanenko, {\it Advances in the Casimir Effect}
(Oxford University Press, Oxford, 2009).
\bibitem{30}
V.~B.~Bezerra, G.~L.~Klimchitskaya,
 V.~M.~Mostepanenko, C.~Romero,
Phys. Rev. D {\bf 81}, 055003 (2010).
\bibitem{31}
V.~B.~Bezerra, G.~L.~Klimchitskaya,
 V.~M.~Mostepanenko, C.~Romero,
Phys. Rev. D {\bf 83}, 075004 (2011).
\bibitem{32}
G.~L.~Klimchitskaya, U.~Mohideen,
V.\ M.\ Mos\-te\-pa\-nen\-ko,
Phys. Rev. D {\bf 86}, 065025 (2012).
\bibitem{33}
V.~M.~Mostepanenko, V.~B.~Bezerra, G.~L.~Klimchitskaya,
C.~Romero,
Int. J. Mod. Phys. A {\bf 27}, 1260015 (2012).
\bibitem{34}
G.~L.~Klimchitskaya, U.~Mohideen,
V.\ M.\ Mos\-te\-pa\-nen\-ko,
Phys. Rev. D {\bf 87}, 125031 (2013).
\bibitem{35}
G.~L.~Klimchitskaya,
V.\ M.\ Mos\-te\-pa\-nen\-ko,
Grav. Cosmol. {\bf 20}, 3 (2014).
\bibitem{36}
R.~S.~Decca, D.~L\'opez, E.~Fischbach, G.~L.~Klimchitskaya,
 D.~E.~Krause, V.~M.~Mostepanenko,
Eur. Phys. J. C {\bf 51}, 963 (2007).
\bibitem{37}
R.~S.~Decca, D.~L\'opez, E.~Fischbach, G.~L.~Klimchitskaya,
 D.~E.~Krause, V.~M.~Mostepanenko,
Phys. Rev. D {\bf 75}, 077101 (2007).
\bibitem{38}
C.~D.~Fosco, F.~C.~Lombardo, F.~D.~Mazzitelli,
Phys. Rev. D {\bf 84}, 105031 (2011).
\bibitem{38a}
L.~P.~Teo, M.~Bordag, V.~Nikolaev,
Phys. Rev. D {\bf 84}, 125037 (2011).
\bibitem{39}
G.~Bimonte, T.~Emig, R.~L.~Jaffe,  M.~Kardar,
Europhys. Lett. {\bf 97}, 50001 (2012).
\bibitem{40}
G.~Bimonte, T.~Emig, M.~Kardar,
Appl. Phys. Lett. {\bf 100}, 074110 (2012).
\bibitem{41}
L.~P.~Teo,
Phys. Rev. D {\bf 88}, 045019 (2013).
\bibitem{42}
S.~D.~Drell, K.~Huang,
Phys. Rev. {\bf 91}, 1527 (1953).
\bibitem{43}
F.~Ferrer, M.~Nowakowski,
Phys. Rev. D {\bf 59}, 075009 (1999).
\bibitem{44}
I.~S.~Gradshtein, I.~M.~Ryzhik,
{\it Table of Integrals, Series and Products}
(Academic Press, New York, 1980).
\bibitem{45}
R.~S.~Decca, E.~Fischbach, G.~L.~Klimchitskaya,
 D.~E.~Krause, D.~L\'opez, V.~M.~Mostepanenko,
Phys. Rev. D {\bf 79}, 124021 (2009).
\bibitem{46}
E.~Fischbach, G.~L.~Klimchitskaya,
 D.~E.~Krause, V.~M.~Mostepanenko,
Eur. Phys. J. C {\bf 68}, 223 (2010).
\bibitem{47}
E.~M.~Lifshitz, L.~P.~Pitaevskii,
{\it Statistical Physics}, Part II
(Pergamon, Oxford, 1980).
\bibitem{53a}
G.~Vasilakis, J.~M.~Brown, T.\ R.\ Kornack,
M.\ V.\ Romalis,
Phys. Rev. Lett. {\bf 103}, 261801 (2009).
\bibitem{53b}
D.~J.~Kapner, T.~S.~Cook, E.~G.~Adelberger,
J.\ H.\ Gundlach, B.\ R.\ Heckel, C.\ D.\ Hoyle,
H.\ E.\ Swanson,
Phys. Rev. Lett. {\bf 98}, 021101 (2007).
\bibitem{53c}
 E.~G.~Adelberger,
B.\ R.\ Heckel, S.\ Hoedl, C.\ D.\ Hoyle,
D.~J.~Kapner, A.\ Upadhye,
Phys. Rev. Lett. {\bf 98}, 131104 (2007).
\bibitem{53d}
A.~V.~Derbin, A.~L.~Frolov, L.~A.~Mitropol'sky,
V.\ N.\ Muratova, D.\ A.\ Semenov, E.\ V.\ Unzhakov,
Eur. Phys. J. C {\bf 62}, 755 (2009).
\bibitem{53e}
A.~V.~Derbin,
V.\ N.\ Muratova, D.\ A.\ Semenov, E.\ V.\ Unzhakov,
Phys. Atom. Nucl. {\bf 74}, 596 (2011).
\bibitem{53f}
J.~Engel, D.~Seckel, A.~C.~Hayes,
Phys. Rev. Lett. {\bf 65}, 960 (1990).
\bibitem{53g}
W.~C.~Haxton, K.~Y.~Lee,
Phys. Rev. Lett. {\bf 66}, 2557 (1991).
\bibitem{53h}
G.~Raffelt,
Phys. Rev. D {\bf 86}, 015001 (2012).
\bibitem{53i}
J.~E.~Moody, F.~Wilczek,
Phys. Rev. D {\bf 30}, 130 (1984).
\bibitem{53j}
V.~V.~Voronin, V.~V.~Fedorov, I.~A.~Kuznetsov,
JETP Lett. {\bf 90}, 5 (2009).
\bibitem{49}
D.~Garcia-Sanches, K.~Y.~Fong, H.~Bhaskaran, S.~Lamoreaux,
H.~X.~Tang,
Phys. Rev. Lett. {\bf 109}, 027202 (2012).
\bibitem{50}
M.~Bordag, G.~L.~Klimchitskaya, V.~M.~Mostepanenko,
Phys. Rev. Lett. {\bf 109}, 199701 (2012).
\end{thebibliography}
\end{document}